\documentstyle[epsfig,10pt]{article}
\begin{document}
\title{\bf Loop Quantum Gravity and Black Hole Physics}
\author{{Carlo Rovelli\footnote{E-mail: rovelli@pitt.edu}}\\ 
	{\it Department of Physics and Astronomy,} \\ 
   {\it University of Pittsburgh, Pittsburgh Pa 15260, USA}}
\date{\today}
\maketitle

\vspace{.2cm}
\begin{abstract}
 \noindent I summarize the basic ideas and formalism of loop quantum
gravity.  I illustrate the results on the discrete aspects of quantum
geometry and two applications of these results to black hole physics.  In
particular, I discuss in detail the derivation of the Bekenstein-Hawking
formula for the entropy of a black hole from first principles. 
 \end{abstract}

\section{Introduction}

\noindent The lack of understanding of the quantum behavior of the
gravitational field, and therefore the lack of understanding of quantum
geometry, remains a major open problem in fundamental physics.  There are
tentative theories which are presently intensively investigated.  For
instance, there is perturbative string theory, and non-perturbative string
theory, much studied in these days.\footnote{For an overview of other
current ideas on quantum geometry, see \cite{jmp95}, \cite{Isham96}, and
\cite{Baez94b}} A less ambitious attempt to solve the problem is
nonperturbative quantum gravity, or ``loop quantum gravity''.\footnote{Ted 
Jacobson calls it ``Loopy quantum gravity''.} This is the project
of taking the conceptual novelties introduced in physics by general
relativity very seriously, and trying to make sense of quantum general
relativity (or any other high energy extension of general relativity)
nonperturbatively. 

The program is based on the hypothesis that perturbative approaches fail
to describe Planck scale physics because at the Planck scale the
separation between a background metric and a quantum field is not
physically justified.  At short scale, spacetime is not Minkoskian.  A
posteriori, loop quantum gravity supports the hypothesis, since the short
structure of the geometry turns out to be strongly non-Minkoskian in the
nonperturbative theory \cite{weave}.  Thus, {\it ``Man shall not separate
what Einstein put together''}: we should not separate the gravitational
field from the metric. 

If this idea is correct, then the quantum gravitational field cannot be
described as a quantum field over a metric manifold, because there is no
background field to provide the metric structure.  Therefore we have to
learn how to construct a quantum field theory living over a structure
weaker than a metric manifold: namely over a differential manifold.  The
aim of nonperturbative quantum gravity is to understand what is quantum
field theory (QFT) on differential manifolds (as opposed to QFT on metric
spaces).  In other words, we want a QFT which is formulated in a
diffeomorphism invariant way, and therefore may incorporate the essential
physical content of general relativity (GR), which --I am convinced-- is
encoded in its active diffemorphism invariance. 

In such a context, most of the techniques of QFT that we like and love
become useless.  Therefore a QFT on a manifold turns out to have a
structure profoundly different from usual QFT's.  The manifold itself is
``washed away'' by diffeomorphism invariance, and therefore the
excitations of the quantum field do not live ``over a space'';  they live
``nowhere'', since they {\it are\ } the space over which physics happens. 
In other words, QFT must undergo the same deep transformation that
classical field theory had to undergo in the evolution from pre-general
relativistic physics to general relativistic physics:  In general
relativistic physics, the ``location'' of physical objects and physical
fields is not determined with respect to a preexisting space.  Rather,
physical quantities (which include the gravitational field) are only
``located'' with respect to each other \cite{location}. The challenge of
quantum gravity is to incorporate this relational notion of localization,
introduced by general relativity, into QFT.  As we shall see, the physical
elementary excitations of the quantum gravitational field are described in
loop quantum gravity by abstract objects (s-knots), which do not live
``inside a given space''.  Rather, they {\it are\ } physical space, at the
quantum level. 

The resulting theory fails to satisfy even the simplest assumptions on
which conventional local QFT is based.  The theory is hard to analyze; it
often contradicts our intuition and some of our accepted believes,
developed in the context of local QFT's.  This is a fact which
unfortunately often complicates the communication between quantum gravity
and other sectors of theoretical physics. 

The idea of exploring quantum GR nonperturbatively is old.  But during the
last decade, the research program has developed intensely, prompted by two
technical advances.  One is the reformulation of classical general
relativity due to Ashtekar \cite{Ashtekar91}, which has substantially
simplified the formalism.  The second is the introduction of the loop
representation for quantum gravity
\cite{Rovelli88,Rovelli90}.\footnote{For a recent overview of canonical
gravity, see \cite{Ehlers94}; for introductions to loop quantum gravity,
see \cite{Rovelli91,Ashtekar92,Smolin93,Gambini,Bruegmann,Ezawa,dpr}.}
The loop representation is a technique for defining a nonperturbative
quantum theory on a manifold.  The idea is to replace creation and
annihilation operators, which are the essential ingredients of
conventional QFT and which make sense only if there is a background
metric, with a different set of operators: the loop operators.  The loop
operators do not require a background metric to be defined.  The theory
defined by a representation of the creation and annihilation operator
algebra is naturally in the Fock, or particle, basis.  The theory defined
by the loop operators is naturally in a basis, denoted the loop basis, or
more precisely the spin network basis, which turns out to be particularly
suitable for dealing with the quantum kinematics and dynamics of the
theory, as well as for analising the nonperturbative aspects of quantum
geometry. 

The loop approach to nonperturbative quantum gravity has now developed in
many directions.  Here, I present a brief overview of the main ideas,
techniques and results, and I focus on a particularly interesting result:
the explicit computation of the spectrum of the area
\cite{area,area2,area3,area4}.  

A traditional problem in quantum gravity is the difficulty of testing
ideas and results \cite{Isham96}. This is due not only to the lack of 
direct experimental
or observational access to Planck scale physics; but also to the intrinsic
difficulty of extracting results from the theory that could be tested
indirectly.  However, there is an area of theoretical physics that gives
us indirect information on quantum gravity: black hole thermodynamics. The
great power of thermodynamics to put constraints on theoretical
constructions, and even provide precise quantitative indications on
microscopic theories is well known: quantum mechanics itself was born to a
large extent in order to satisfy thermodynamical consistency requirements
(Planck's spectrum, solid state...). Now, black hole thermodynamics
derives a surprising set of simple laws just from classical general
relativity and quantum field theory in curved spacetime (for an
introduction, see \cite{wald}). These laws have not been experimentally
tested, but are very well motivated.  However, they are thermodynamical
``phenomenological'' laws, and their derivation from first principles
requires a quantum theory of gravity, and, at present, is lacking. 

This state of affairs provides the ideal testing ground for loop quantum
gravity. The study of the applications of loop quantum gravity to black
hole thermodynamics has just begun.  Here, I describe two of these
applications. The first \cite{Barreira} is a discussion of the Bekenstein
Mukhanov effect \cite{bm}. The second \cite{Rovelli96,kirill} and more
important application is a derivation of the Bekenstein-Hawking black hole
entropy formula \cite{bek,haw} from first principles.  In some parts of
this lecture I will follow, and sometimes expand, references
\cite{dpr,Barreira,Rovelli96}. 

\section{Overview of loop quantum gravity}

Classical general relativity can be formulated in phase space
form as follows \cite{Ashtekar91,Barbero,Thiemann}.  We fix a
three-dimensional manifold $M$ and consider two real (smooth) $SO(3)$
fields $A_a^i(x)$ and $\tilde{E}^a_i(x)$ on~$M$.  We use
$a,b,\ldots=1,2,3$ for (abstract) spatial indices and $i,j,\ldots=1,2,3$
for internal $SO(3)$ indices.  We indicate coordinates on $M$ with~$x$. 
The relation between these fields and conventional metric gravitational
variables is as follows:  $\tilde{E}^a_i(x)$ is the (densitized) inverse
triad, related to the three-dimensional metric $g_{ab}(x)$ of the
constant-time surface by %
 \begin{eqnarray} 
 g\ g^{ab} =
 \tilde{E}^a_i\tilde{E}^b_i, 
 \end{eqnarray} %
 where $g$ is the determinant
of $g_{ab}$; and %
\begin{equation} A_a^i(x)=\Gamma_a^i(x)+k_a^i(x); 
\label{real} 
\end{equation} 
Where $\Gamma_a^i(x)$ is the $SU(2)$ spin
connection associated to the triad and $k_a^i(x)$ is the extrinsic
curvature of the three surface (up to indices' position).  Notice the
absence of the $i$ in (\ref{real}), which yields the {\it real \/}
Ashtekar connection. 

The spinorial version of the Ashtekar variables is given in terms of 
the Pauli matrices
${\sigma_i}, i=1,2,3$, or the $su(2)$ generators
${\tau_i} = -  \frac{{\rm i}}{2} \ {\sigma_i}$, by
\begin{eqnarray}
\tilde{E}^{a}(x) 
        &=& - {\rm i}\ \tilde{E}^a_i(x) \ {\sigma_i}
         =  2 \tilde{E}^a_i(x) \ {\tau_i} \\
A_{a}(x)&=& - \frac{{\rm i}}{{2}}\ {A}_a^i(x)\ {\sigma_i}
         = {A}_a^i(x)\ {\tau_i}\,.\,.
\end{eqnarray}
$A_a(x)$ and $\tilde{E}^a(x)$ are $2\times 2$
complex matrices.  

The theory is invariant under local $SO(3)$ gauge, 
three-dimensional diffeomorphisms of the manifold on which the fields are
defined, as well as under (coordinate) time translations generated by the
Lorentzian Hamiltonian constraint. The full dynamical content of GR is
captured by the three constraints that generate these gauge invariances
\cite{Ashtekar91}.  The Lorentzian Hamiltonian constraint does not have a 
simple polynomial form if we use the real connection (\ref{real}). For 
a while, this fact was considered an obstacle defining the 
quantum Hamiltonian constraint; therefore the complex version of the 
connection was mostly used.  However, Thiemann has recently   
succeded in constructing a satisfactory Lorentzian quantum hamiltonian 
constraint \cite{Thomas} in spite of the non-polynomiality of the 
classical expression.  This is the reason we use here the real 
connection. This choice has the  advantage of greatly simplifying the 
``reality conditions'' problem. 
 
To construct the quantum theory, we have to promote the fields to
operators on a Hilbert space.  One possibility is to consider the positive
and negative frequencies of A and E, and define a Fock representation. 
The definition of positive and negative frequencies requires a metric. 
Thus, one may consider an unperturbed background field around which
expanding A and E, and use the unperturbed field as background metric. The
problem is that the expansion becomes unsuitable precisely at the Planck
scale, which is the scale we are interested in. 

The loop representation is based on the choice of other quantities to be
promoted as basic operators. These are: the trace of the holonomy of the
Ashtekar connection, which is labeled by loops on the three manifold; and
the higher order loop variables, obtained inserting the $E$ field (in $n$
distinct points, or ``hands'' of the loop variable) into the holonomy
trace.  More precisely, given a loop $\alpha$ and the points
$s_1,s_2,\ldots,s_n\in\alpha$ we define: 
 \begin{eqnarray}
  {\cal T}[\alpha]          &=& - {\rm Tr} [U_\alpha] ,
\label{t0}  \\
  {\cal T}^a[\alpha](s)     
      &=& - {\rm Tr} [U_\alpha(s,s) \tilde{E}^a(s)] 
\end{eqnarray}
and, in general
\begin{eqnarray}
{\cal T}^{a_1a_2}[\alpha](s_1,s_2) &=& 
  - {\rm Tr} [U_\alpha(s_1,s_2) \tilde{E}^{a_2}(s_2) 
                U_\alpha(s_2,s_1) \tilde{E}^{a_1}(s_1) ] ,
\\
{\cal T}^{a_1\ldots a_N}[\alpha](s_1 \ldots s_N) 
&=&  - {\rm Tr} [U_\alpha(s_1,s_N) \tilde{E}^{a_N}(s_N)
                U_\alpha(s_N,s_{N-1}) \ldots 
\tilde{E}^{a_1}(s_1) ]  
\nonumber
\end{eqnarray}
 where $U$ is the parallel propagator of $A_a$ along $\gamma$. (See
\cite{dpr} for more details.) These are the loop observables.  They
coordinatize the phase space and have a closed Poisson algebra. Thus, we
may pick a unitary representation of this algebra as the definition of the
kinematic of the quantum theory. 

\subsection{The Hilbert space}

A representations of the loop algebra can be defined as follows (The first
introduction of the loop representation, in the context of Yang Mills
theory, is in \cite{Gambini81}).  We consider the free algebra 
${\cal A}^f[{\cal L}]$ over the set of the loops in the three manifold, 
namely
the set of objects $\Phi$ which are (finite) formal linear combinations of
formal products of loops: 
 \begin{equation}
\Phi = c_0 + \sum_i c_i ~[\alpha_i] + \sum_{jk} c_{jk}                 
	~[\alpha_j][\alpha_k]~ + \ldots ~~, \label{fla}
\end{equation}
 where the $c$'s are arbitrary complex number and the
$\alpha$'s are loops (see also \cite{Ashtekar92}).
The loop observable (\ref{t0}) has an immediate extension to 
this algebra as\footnote{The following 
formula corresponds to equation (2.15) in ref.\cite{dpr}.  However, equation 
(2.15) in \cite{dpr} contains an additional $(-2)$ factor in the first term. 
The $(-2)$, and the motivation given in \cite{dpr} for its introduction, 
are not correct. I thank Laszlo Szabados for pointing this 
out.} 
\begin{equation}
	{\cal T}[\Phi]= c_0 + \sum_i c_i {\cal T}[\alpha_i] + \sum_{jk} 
c_{jk} {\cal T}[\alpha_j]\,{\cal T}[\alpha_k]~ + \ldots ~~\ .
\end{equation}
The algebra ${\cal A}^f[{\cal L}]$ contains the ideal
\begin{equation}
{\cal K} = \{\Phi\in {\cal A}^f[{\cal L}]\ \ |\ \ {\cal T}[\Phi] 
=0\},
\end{equation}
and we define the carrier space $\cal V\ $ of the
representation by
\begin{equation}
{\cal V}= {\cal A}^f[{\cal L}]/{\cal K}.
\end{equation}
In other words, the state space of the loop representation
is defined as the space of the equivalence classes of
linear combinations of multiloops, under the equivalence
defined by the Mandelstam relations
\begin{equation}
  \Phi\sim\Psi \ \ {\rm if} \ \ {\cal T}[\Phi] ={\cal
T}[\Psi],
\label{eq:Mandelstam}
\end{equation}
namely by the equality of the corresponding holonomies
\cite{Ashtekar92}. 

There is natural basis in this linear space, denoted the spin network
basis, which was introduced in \cite{spinnet}, and developed in
\cite{Baez95a}.  This is defined as follows.  A spin network $S$ is here a
graph imbedded in the three dimensional space $M$, with a ``color'' (a
positive integer) assigned to each link of the graph. Vertices with
valence higher than three are (arbitrarily) expanded in tree-like
``virtual'' trivalent graphs and the ```virtual'' edges are colored as
well (see \cite{dpr}). Colors satisfy a condition at the vertices: in a
trivalent vertex, each color is not larger than the sum of the other two
(Clebsh-Gordon condition), and the sum of the three colors is even. 

There exists a procedure to associate a linear combination of formal
products of loops, and therefore an element of the quantum state space $\cal
V$, to each such spin network.  The procedure (introduced by Penrose
\cite{Penrose}) consists in replacing each (real and virtual) edge colored
$p$ with $p$ overlapping lines, joining these lines at the vertices and
then anti-symmetrizing the lines in each (real and virtual) edge. One can
then prove that the quantum states $|S\rangle$ obtained in this way form a
basis in $\cal V$. 

Finally, a scalar product is naturally defined over $\cal V$ (see 
\cite{dpr}, and below). We can complete in the Hilbert norm, 
obtaining the (``unconstrained'' or ``kinematical'')  Hilbert space of 
the quantum theory, which we denote as $\cal H$. 

\subsection{Structures in $\cal H$}

The Hilbert space $\cal H$ has a rich structure that has been extensively
explored.  First of all, the spin network states satisfy the Kauffman
axioms of the tangle theoretical version of recoupling theory
\cite{Kauffman94} (in the ``classical'' case $A=-1$) at all the points
(in 3d space) in which they meet \footnote{This fact is often
misunderstood: recoupling theory lives in 2d and is associated by Kauffman
to knot theory by means of the usual projection of knots from 3d to 2d.
Here, the Kauffmann axioms are not satisfied at the intersections created
by the 2d projection of the spin network, but only at the true
intersections in 3d. See \cite{dpr} for a detailed discussion.}.  For 
instance, weave
consider a 4-valent intersection of four edges colored $a, 
b, c, d$. The color of the vertex is determined by expanding the 
4-valent intersection into a trivalent tree; in this case, we have a 
single internal edge. The expansion can be done in different ways (by 
pairing edges differently). These are related to each other by the 
recoupling theorem of pg.\ 60 in Ref.\ \cite{Kauffman94}
\begin{equation}
\begin{array}{c}\setlength{\unitlength}{1 pt}
\begin{picture}(50,40)
          \put( 0,0){$a$}\put( 0,30){$b$}
          \put(45,0){$d$}\put(45,30){$c$}
          \put(10,10){\line(1,1){10}}\put(10,30){\line(1,-1){10}}
          \put(30,20){\line(1,1){10}}\put(30,20){\line(1,-1){10}}
           \put(20,20){\line(1,0){10}}\put(22,25){$j$}
          \put(20,20){\circle*{3}}\put(30,20){\circle*{3}}
\end{picture}\end{array}
    = \sum_i  \left\{\begin{array}{ccc}
                      a  & b & i \\
                      c  & d & j
              \end{array}\right\}
\begin{array}{c}\setlength{\unitlength}{1 pt}
\begin{picture}(40,40)
      \put( 0,0){$a$}\put( 0,40){$b$}
      \put(35,0){$d$}\put(35,40){$c$}
     \put(10,10){\line(1,1){10}}\put(10,40){\line(1,-1){10}}
      \put(20,30){\line(1,1){10}}\put(20,20){\line(1,-1){10}}
      \put(20,20){\line(0,1){10}}\put(22,22){$i$}
      \put(20,20){\circle*{3}}\put(20,30){\circle*{3}}
\end{picture}\end{array}
\label{rec}
\end{equation}
where the quantities 
$\left\{\begin{array}{ccc}
a & b & i \\ c & d & j  \end{array}\right\}$ 
are $su(2)$
six-j symbols (normalized as in \cite{Kauffman94}). Equation 
(\ref{rec}) follows just from the definitions given above. Recoupling 
theory provides a powerful computational tool in this context. 

Since spin network states satisfy recoupling theory, they form a
Temperley-Lieb algebra \cite{Kauffman94}.  The scalar product in $\cal H$
is given by the Temperley-Lieb trace of the spin networks, or, equivalently
by the Kauffman brackets, or, equivalently, by the chromatic evaluation of
the spin network. Spin network states form an orthogonal base. See 
Ref.~\cite{dpr} for an extensive discussion of these relations. 

Next, the space $\cal H$ can be constructed as the projective limit of a
(projective) family of Hilbert spaces ${\cal H}_\Gamma$ of $SU(2)$ lattice
gauge theories defined over arbitrary lattices $\Gamma$ in three-space
\cite{Ashtekar95}.  The space ${\cal H}_\Gamma$ naturally sits into the
space ${\cal H}_{\Gamma'}$ when the graph $\Gamma$ is a subgraph of
$\Gamma'$, and, correspondingly, the spaces ${\cal H}_\Gamma$ form a
projective family. 

Next, $\cal H$ can be viewed as the space of gauge-invariant functions
over (the closure in a suitable norm of) the space $\cal A$ of the $SU(2)$
gauge connections, which are square integrable under the
Ashtekar-Lewandowski-Baez measure $d\mu_{ALB}[A]$ \cite{Lewandowski}.
$\cal A$ can be thought as a space of ``distributional connections''.  The
Ashtekar-Lewandowski-Baez measure is a diffeomorphism invariant measure
over such space. (Or, equivalently, a ``generalized measure'' over the
space of smooth connection \cite{Baez95a}.) The cylindrical functions
over which the measure is constructed correspond precisely to the
spin network states defined above.  

The relation is as follows.  When restricted to the (dense) subspace of
$\cal A$ formed by smooth connections, the cylindrical function
$\psi_S[A]=\langle A| S\rangle$ corresponding to a given spin network state
$|S\rangle$ is formed by parallel propagators of the $SU(2)$ connection
along the edges of $S$, in the representation $p/2$, where $p$ is the
color of the edge, contracted at the vertices by means of invariant
tensors in the tensor product of the representations associated to the
edges joining at the vertex. The colors of the vertex (namely the colors
of the internal edges) label the independent invariant
tensors\footnote{Because a basis of invariant $SU(2)$ tensor on the tensor
product of a finite number of irreps. is obtained by progressively
decomposing tensor products or irreps. into irreps., two by two.}. This
construction gives a rigorous meaning to the loop transform, which was
used as an heuristic devise to build the loop representation in
\cite{Rovelli90}. In fact, we can write, for every spin netwok $s$, and
every state $\psi[A]$ 
  \begin{equation}
	\psi(S)=\langle S|\psi\rangle=\int d\mu_{ALB}[A]\ 
	\bar\psi_S[A]\ \psi[A]
  \end{equation}
One can show that this equation defines a unitary mapping between the two 
presentations of $\cal H$: the ``loop representation'', in  which one 
works in terms of the basis $|S\rangle$; and the ``connection 
representation'', in which one uses wave functionals $\psi[A]$.

For a recent discussion of the unitary equivalence between loop and
connection representations see \cite{Lewandowski96} and \cite{DePietri96}. 
The relation between the two representations is also an implementation of
the well known duality between $SU(2)$ representation theory and the
combinatorics of planar loops. This duality has been much exploited in
physical applications, and underlies all graphical methods for dealing
with $SU(2)$ representation theory \cite{Brink68}.  It was Penrose who
first had the intuition that this mathematics could be relevant for
describing the quantum properties of the geometry, and who gave the first
version of spin network theory \cite{Penrose}. 

Finally, Ashtekar and Isham \cite{AshtekarIsham} have recovered the
representation of the loop algebra by using C*-algebra representation
theory: The space ${\cal A}/{\cal G}$, where $\cal G$ is the group of 
local $SU(2)$ transformations, is precisely the Guelfand spectrum of the
abelian part of the loop algebra. One can show that this is a suitable
norm closure of the space of smooth $SU(2)$ connections over physical
space, modulo gauge transformations. 

Thus, a number of powerful mathematical tools are at hand for dealing with
nonperturbative quantum gravity. Some of these have already been
extensively used in this context. These include:  Penrose's spin network
theory, $SU(2)$ representation theory, Kauffman tangle theoretical
recoupling theory, Temperly-Liebb algebras, Gelfand's $C^*$algebra
spectral representation theory, infinite dimensional measure theory and
differential geometry over infinite dimensional spaces. 

\subsection{The representation}

We now define the quantum operators, corresponding to the
$\cal T$-variables, as linear operators on $\cal H$. These
form a representation of the loop variables Poisson algebra.
The operator ${\hat{\cal T}}[\alpha]$, acting on a state
${\left\langle{\Phi}\right|}$ simply adds a loop to
${\left\langle{{\Phi}}\right|}$:
\begin{eqnarray}
 \bigg\langle {c_0 + \sum_i c_i ~[\alpha_i]
  + \sum_{ij} c_{ij} ~[\alpha_i][\alpha_j]~
  + \ldots}\bigg|\ \  {\hat{\cal T}}[\alpha] =&&
\nonumber \\
= \bigg\langle {c_0 [\alpha]+ \sum_i c_i ~[\alpha_i] [\alpha]
  + \sum_{ij} c_{ij} ~[\alpha_i][\alpha_j] [\alpha]
  + \ldots}\bigg|\ &.& 
 \end{eqnarray} 
 (The consuetudinal bra notation is just a historical left-over from the 
period 
when the scalar product was not known.) Higher order loop operators are
expressed in terms of the elementary ``grasp'' operation: acting on an
edge with color $p$, the hand of the loop operator creates two ``virtual''
trivalent vertices, one on the spin-network state and one the loop of the
operator.  The two virtual vertices are joined by a virtual edge of color
2.  
\begin{equation} 
\begin{array}{c}
 \mbox{\epsfig{file=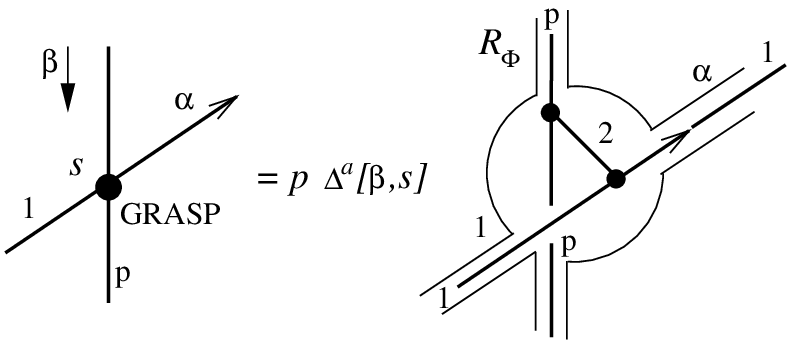}}
 \end{array}
\end{equation}
where we have introduced the elementary length $l_0$ by
\begin{equation}
 l^2_0 = \hbar G = \frac{16 \pi \hbar G_{{\rm Newton}}}{c^3} 
 = 16 \pi\ l^2_{Planck}
\end{equation}
and 
 \begin{equation}
 \Delta^a[\beta,s] = \int_\beta d\tau ~\dot{\beta}^a(\tau) 
                   \delta^3 [\beta(\tau),s]. \label{delta} 
 \end{equation}
The sign of the tangent of $\beta$ in $\Delta^a[\beta,s]$ is determined by
the orientation of $\beta$ consistent with the positive-terms of the loop
expansion of the spin network.  Higher order loop operators act 
similarly.  One can verify that these operators provide a representation 
of the classical Poisson loop algebra. 

All the operators in the theory are then constructed in terms of these
basics loop operators, in the same way in which in conventional QFT one
constructs all operators, including the Hamiltonian, in terms of creation and
annihilation operators. The construction of the composite operators
requires the development of regularization techniques that can be used in
the absence of a background metric. These have been introduced in
\cite{Smolin93} and developed in \cite{weave,area,dpr,Ashtekar95}. I will
illustrate these techniques below.

\subsection{Diffeomorphism invariance}

The next step in the construction of the theory is to factor away
diffeomorphism invariance. This is a key step for two reasons.  First of
all, $\cal H$ is a ``huge'' non separable space. It is far ``too large''
for a quantum field theory.  However, most of this redundancy is all
gauge, and disappears when one solves the diffeomorphism constraint,
defining the physical Hilbert space ${\cal H}_{Ph}$.  This is the reason
for which the loop representation, as defined here, is of great value 
in diffeomorphism invariant theories only. 

The second reason is that ${\cal H}_{Ph}$ turns out to have a natural
basis labeled by knots. More precisely by ``s-knots''. An s-knot $s$ is
an equivalence classes of spin networks $S$ under diffeomorphisms.  An
s-knot is characterized by its ``abstract'' graph (defined only by the
adjecency relations between edges and vertices), by the coloring, and by
its knotting and linking properties, as in knot-theory.\footnote{Finite
dimensional moduli spaces associated with high valence intersections
appear \cite{norbert}. Their physical relevance is unclear at this stage.}
Thus, the physical quantum states of the gravitational field turn out to
be essentially classified by knot theory. 

There are various equivalent way of obtaining ${\cal H}_{Ph}$ from ${\cal
H}$. One can use regularization techniques for defining the quantum
operator corresponding to the classical diffeomorphism constraint in terms
of elementary loop operators, and then find the kernel of such operator. 
Equivalently, one can factor ${\cal H}$ by the natural action of the
Diffeomorphism group that it carries. Namely \begin{equation} {\cal
H}_{Ph}={{\cal H}\over Diff(M)}.  \end{equation} For a rigorous way for
defining such a quotient of an Hilbert space by an infinite dimensional
group, see \cite{Ashtekar95} and references therein. 

\subsection{Dynamics}

Finally, the definition of the theory is completed by giving the
Hamiltonian constraint.  A number of approaches to the definition of a
Hamiltonian constraint have been attempted in the past, with various
degrees of success.  Recently, however, Thiemann has succeded in providing
a regularization of the Hamiltonian constraint that yields a well defined,
finite operator in ${\cal H}_{Ph}$.  Thiemann's construction \cite{Thomas}
is based on several clever ideas. I will not describe it here.  Rather, I
will sketch below the final form of the constraint (for the Lapse=1 case), 
following \cite{jmp}. 

I begin with the Euclidean Hamiltonian constraint $H_E$. We have 
 \begin{equation}
	\hat H |s \rangle = \sum_i \sum_{(IJ)}\ 
\sum_{\epsilon=\pm 1} \ 
	\sum_{\epsilon'=\pm 1}\   
	A_{\epsilon\epsilon'}(p_i...p_n)
	 \ \hat D_{i;(IJ),\epsilon\epsilon'}\  |s>. 
\label{lee}
\end{equation}
Here $i$ labels the vertices of the s-knot $s$; 
$(IJ)$ labels couples of (distinct) edges emerging from $i$. 
$p_1...p_n$ are the colors the edges emerging from $i$.   $\hat 
D_{i;(IJ)\epsilon\epsilon'}$ is the operator that acts on an $s\,$-knot 
by: (i) creating two additional vertices, one along each of the two 
links $I$ and $J$; (ii) creating a novel 
link, colored 1, joining these two nodes, (iii) assigning the coloring 
$p_I+\epsilon$ and, respectively, $q_J+\epsilon'$ to the links that 
join the new formed nodes with the node $i$. This is illustrated in 
Figure 1. 
\vskip.5cm
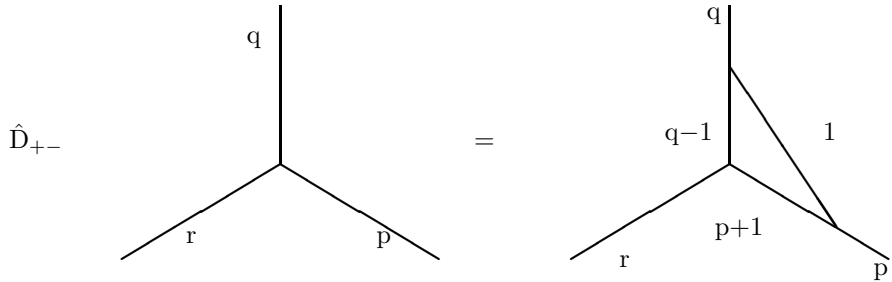
\begin{figure}
\begin{picture}(344,114)(105,606)
\thicklines
\put(201,648){\line(-5,-3){ 60}}
\put(201,648){\line( 5,-3){ 60}}
\put(201,648){\line( 0, 1){ 60}}
\put(162,618){\makebox(0,0)[lb]{\raisebox{0pt}[0pt][0pt]{
 r}}}
\put(185,694){\makebox(0,0)[lb]{\raisebox{0pt}[0pt][0pt]{
 q}}}
\put(234,618){\makebox(0,0)[lb]{\raisebox{0pt}[0pt][0pt]{
 p}}}
\put(95,654){\makebox(0,0)[lb]{\raisebox{0pt}[0pt][0pt]{
\^D${}_{+-}$}}}
\put(371,685){\line( 2,-3){ 40.462}}
\put(371,648){\line(-5,-3){ 60}}
\put(371,648){\line( 5,-3){ 60}}
\put(371,648){\line( 0, 1){ 60}}
\put(271,654){\makebox(0,0)[lb]{\raisebox{0pt}[0pt][0pt]{
=}}}
\put(326,609){\makebox(0,0)[lb]{\raisebox{0pt}[0pt][0pt]{
 r}}}
\put(422,606){\makebox(0,0)[lb]{\raisebox{0pt}[0pt][0pt]{
 p}}}
\put(359,703){\makebox(0,0)[lb]{\raisebox{0pt}[0pt][0pt]{
 q}}}
\put(403,657){\makebox(0,0)[lb]{\raisebox{0pt}[0pt][0pt]{
 1}}}
\put(343,657){\makebox(0,0)[lb]{\raisebox{0pt}[0pt][0pt]{
q$-$1}}}
\put(362,621){\makebox(0,0)[lb]{\raisebox{0pt}[0pt][0pt]{
p$+$1}}}
\end{picture}
\caption{Action of $\hat D_{i;(IJ)\epsilon\epsilon'}$.}
\end{figure}

The coefficients $A_{\epsilon\epsilon'}(p_i...p_n)$, which are finite, can
be expressed explicitly (but in a rather laborious way) in terms of
products of linear combinations of $6-j$ symbols of $SU(2)$, following the
techniques developed in detail in \cite{dpr}. Some of these coefficients 
have been explicitly computed \cite{a}. The Lorentzian Hamiltonian
constraint is given by a similar expression, but quadratic in the $\hat D$
operators. 

\subsection{Developments}

In the previous section, I have sketched the basic structure of the loop
representation.  This has been developed in a great number of directions. 
Without any ambition of completeness, I list below some of these
developments. 

\begin{itemize}

\item {\it Solutions of the Hamiltonian constraints. } One of the most
surprising results of the theory is that it has been possible to find
exact solutions of the Hamiltonian constraint. This follows from the key
result that the action of the Hamiltonian constraints is non vanishing
only over vertices of the s-knots \cite{Rovelli88,Rovelli90}.  Therefore
s-knots without vertices are physical states that solve the quantum
Einstein dynamics. There is an infinite number of independent states of
this sort, classified by conventional knot theory.  The physical
interpretation of these solutions is still rather obscure. But the issue
has received much attention, and various other solutions have been found. 
See the recent review \cite{Ezawa} and reference therein. See also
\cite{Gambini,Bruegmann,FalseIntersec,rp}. 

\item {\it Time evolution. Strong field perturbation expansion. 
 ``Topological Feynman rules''. } Trying to describe the temporal
evolution of the quantum gravitational field by solving the Hamiltonian
constraint yields the conceptually well-defined \cite{RovelliTime}, but
notoriously very non-transparent frozen-time formalism. An alternative is
to study the evolution of the gravitational degrees of freedom with
respect to some matter variable, coupled to the theory, which plays the
role of a phenomenological ``clock''.  This approach has lead to the
definition of a physical Hamiltonian \cite{Rovelli94b}, and to a
preliminary investigation of the possibility of transition amplitudes
between s-knot states, order by order in a (strong coupling) perturbative
expansion \cite{Rovelli95b}.  In this context, diffeomorphism invariance,
combined with the key result that the Hamiltonian constraint acts on
vertices only, imply that the ``Feynman rules'' of such an expansion are
purely topological and combinatorial. 

\item {\it Classical limit.  Quantum states representing flat spacetime. 
Weaves. Discrete small scale structure of flat space. } The s-knots do not
represent excitations of the quantum gravitational field over flat space,
but rather over a ``no-space'', or $g_{\mu\nu}=0$ solution.  A natural
problem is then how flat space (or any other smooth geometry) is
represented in the theory.  Notice that in a general relativistic context
the Minkowski solution does not have all the properties of the
conventional field theoretical vacuum.  (In gravitational physics there is
no real equivalent of the conventional vacuum, particularly in the
spacially compact case.) One then expects that flat space is represented
by some highly excited state in the theory.  States in $\cal H$ that
describe flat space when probed at low energy (large distance) have been
studied in \cite{weave}.  These have a discrete structure at the Planck
scale.  Furthermore, small excitations around such states have been
considered in \cite{Iwasaki}, where it is shown that $\cal H$ contains all
``free graviton'' physics, in a suitable approximation. 

\item {\it Fermions. } Fermions have been added to the theory
\cite{Rovelli94}. Remarkably, all the important results of the pure GR
case survive in the GR+fermions theory.  Not surprisingly, fermions can be
described as open ends of ``open spin networks''. 

\item  {\it Maxwell. } The extension of the theory to the Maxwell 
field has been studied in \cite{Krasnov95}.

\item {\it Application to other theories. } The loop representation has
been applied in various other contexts such as 2+1 gravity \cite{21}, some
topological field theories, and others. 

\item {\it Lattice and simplicial models. } A number of very interesting
discretized versions of the theory are being studied. See in particular
\cite{discrete}. 
 
\item {\it Spectra of geometrical quantities. Area and volume. } Finally,
the results that I consider most characteristic and potentially most
fruitful regard spectral properties of geometrical quantities, such as
area and volume of regions physically defined (say by matter).  I will
focus on these results in the next section. 

\end{itemize}

\section{Area}

Consider a physical situation in which the gravitational field is
interacting with some matter. We are interested in the area of a surface
defined by the matter. For instance, imagine we are studying the explosion
of a supernova.  One second after the explosion, the matter of the
supernova is approximately spherical, and defines a surface $\Sigma$: the
surface of the star. The physical area of $\Sigma$ depends on the matter
as well as on the metric, namely on the gravitational field.  In a quantum
theory of gravity, the gravitational field is a quantum field operator,
and therefore we must describe the area of $\Sigma$ in terms of a quantum
observables described by an operator $\hat A$.  We now ask what is the
quantum operator $\hat A$ in nonperturbative quantum gravity. 

Consider a 2d surface $\Sigma$ imbedded in $M$ 
with coordinates
$\sigma^u=(\sigma^1,\sigma^2)$.  We
write $S:\Sigma \longrightarrow M, \sigma^u
\longrightarrow ~x^a(\sigma)$. The metric and the normal
one form on $\Sigma$ are given by
\begin{eqnarray}
g^{\Sigma}   &=& S^\star~g,     
\qquad\qquad  g^{\Sigma}_{uv}= 
\frac{\partial x^a}{\partial \sigma^u}
\frac{\partial x^b}{\partial \sigma^v}  g_{ab};  \\
n_a  &=& \frac{1}{2}\epsilon^{uv} \epsilon_{abc} 
\frac{\partial x^b}{\partial \sigma^u}
\frac{\partial x^c}{\partial \sigma^v}.
\end{eqnarray}
The area of $\Sigma$ is 
\begin{eqnarray}
A[\Sigma] &=& \int_\Sigma\!d^2\sigma~\sqrt{\det g^{\Sigma}} 
= \int_\Sigma\!d^2\sigma~\sqrt{\frac{1}{2} 
\epsilon^{u\bar{u}}\epsilon^{\bar{v}\bar{v}} 
g^{\Sigma}_{uv}g^{\Sigma}_{\bar{u}\bar{v}}} 
\nonumber \\
&=& \int_\Sigma\! d^2\sigma
~\sqrt{n_a n_b \tilde{E}^{ai} \tilde{E}^{b}_{i}}, 
\label{area}
\end{eqnarray}
(On the role of played by surface area in the Ashtekar's
formulation of GR, see \cite{Rovelli93b}.)  We want to
construct the quantum area operator $\hat A[\Sigma]$,
namely a function of the loop representation operators whose
classical limit is $A[\Sigma]$.  Following conventional
quantum field theoretical techniques, we deal with
operator products by defining $\hat A[\Sigma]$ as a limit
of regularized operators $\hat A_{\epsilon}[\Sigma]$ that
do not contain operator products.  The difficulty in the
present context is to find a regularization that does not
break general covariance.  This can be achieved by a
geometrical regularization \cite{Smolin93}.

Following \cite{area}, we begin by constructing a
classical regularized expression for the area, namely a
one parameter family of classical functions of the loop
variables $A_\epsilon[\Sigma]$ which converges to the area
as $\epsilon$ approaches zero.\footnote{I simplify here. For 
regularization that works in the general case, see \cite{area3}}  
Consider a small region $\Sigma_\epsilon$ of the surface $\Sigma$, whose
coordinate area goes to zero with $\epsilon^2$.  For every
$s$ in $\Sigma$, the smoothness of the classical fields
implies that $\tilde{E}^a(s) = \tilde{E}^a(x_I) +
O(\epsilon)$, where $x_I$ is an arbitrary fixed point in
$\Sigma_\epsilon$. Also, $U_\alpha(s,t)_A^{~B} =
\delta_A^{~B} + O(\epsilon)$ for any $s,t\in \Sigma_I$
and $\alpha$ a (coordinate straight) segment joining $s$
and $t$.  It follows that to 
zeroth order in $\epsilon$
\begin{eqnarray}
 {\cal T}^{ab}[\alpha_{st}](s,t) 
     &=& - {\rm Tr}\left[
             \tilde{E}^a(s) U_\alpha(s,t)      
             \tilde{E}^b(t) U_\alpha(t,s)      
           \right]
\nonumber \\
     &=& 2 \tilde{E}^{ai}(x_I)\tilde{E}^{b}_i(x_I) .
\label{2Tlimit}
\end{eqnarray}
Using this, we can write  
\begin{eqnarray} 
&& \epsilon^4 \tilde{E}^{ai}(x_I)\tilde{E}^{b}_i(x_I) 
  =\frac{1}{2}\int_{\Sigma_{\epsilon}} d^2\!\sigma~ n_a(\sigma) 
              \int_{\Sigma_{\epsilon}} d^2\!\tau ~  n_b(\tau)
\nonumber \\ && ~~~~~~~~ 
         ~{\cal T}^{ab}[\alpha_{\sigma\tau}](\sigma,\tau) 
	 + O(\epsilon),
\label{ee}
\end{eqnarray}
where $\alpha_{\sigma\tau}$ is, say, a (coordinate) circular
loop with the two points $\sigma$ and $\tau$ on antipodal
points.  Next, consider the area of the full surface
$\Sigma$.  By definition of Riemann integral,
(\ref{area}) can be written as
\begin{eqnarray}
A[\Sigma] &=& \int_\Sigma\! d^2\sigma
~\sqrt{n_a n_b \tilde{E}^{ai} \tilde{E}^{b}_{i}} 
\\&=& 
= \lim_{ \stackrel{{\scriptstyle  N\to\infty}}{
         {\scriptstyle  \epsilon\to 0}}} 
    \sum_{I_\epsilon}  \epsilon^2 
~\sqrt{n_a(x_I) n_b(x_I) 
  \tilde{E}^{ai}(x_I) \tilde{E}^{b}_{i}(x_I)}
\label{riemann} 
\nonumber
\end{eqnarray}
where, following Riemann we have partitioned the surface
$\Sigma$ in $N$ small surfaces $\Sigma_{I_\epsilon}$ of
coordinate area $\epsilon^2$ and $x_I$ is an arbitrary
point in $\Sigma_{I_\epsilon}$.  Inserting
(\ref{ee}) in (\ref{riemann}), we obtain the desired
regularized expression for the classical area, suitable to
be promoted to a quantum loop operator
\begin{eqnarray}
A[\Sigma] &=& \lim_{\epsilon\to 0}\ A_\epsilon[\Sigma] ~~,  
\label{limitc}
\\
A_\epsilon[\Sigma] &=& \sum_{I_\epsilon} \sqrt{A^2_{I_\epsilon}}
~~,\\
A^2_{I_\epsilon} &=& \frac{1}{2}
       \int_{\Sigma_{I_\epsilon}\otimes\Sigma_{I_\epsilon}} 
       \!\!\!\!\!\!\!\!\! d^2\!\sigma  d^2\!\tau ~ 
        n_a(\sigma) n_b(\tau) 
          ~{\cal T}^{ab}[\alpha_{\sigma\tau}](\sigma,\tau). 
\end{eqnarray}
Notice that the powers of the regulator $\epsilon$ in
(\ref{ee}) and (\ref{riemann}) combine nicely, so that
$\epsilon$ appears in (\ref{limitc}) only in the
integration domains.

We are now ready to define the area operator:
\begin{eqnarray}
\hat A[\Sigma] &=& 
\lim_{\epsilon\to 0}\ \hat A_\epsilon[\Sigma],  
\label{limit}
\\
 A_\epsilon[\Sigma] &=&  \sum_{I_\epsilon} 
\sqrt{\hat{A}^2_{I_\epsilon}}, 
\label{root}
\\
\hat{A^2_{I_\epsilon}} &=& \frac{1}{2}
        \int_{\Sigma_{I_\epsilon}\otimes\Sigma_{I_\epsilon}} 
           \!\!\!\!\!\!\!\!\!\!\! d^2\!\sigma  d^2\!\tau ~ 
           n_a(\sigma) n_b(\tau) 
           ~{\hat{\cal T}}^{ab}[\alpha_{\sigma\tau}](\sigma,\tau). 
\label{sigmatau}
\end{eqnarray}
The meaning of the limit in (\ref{limit}) is discussed
in detail in \cite{dpr}. 

We now study the action of the area operator $\hat
A[\Sigma]$ given in (\ref{limit}) on a spin network state
${\left\langle{{S}}\right|}$. We
label by an index $i$ the points where the spin network
graph $\Gamma_S$ and the surface $\Sigma$
intersect.   (Here we disregard spin
networks that have a vertex lying on $\Sigma$ or a
continuous number of intersection points with $\Sigma$. 
The complete spectrum of the area, including these cases is
given in  \cite{area2,area3}.)

For small enough $\epsilon$, each intersection $i$ will lie
inside a distinct $\Sigma_{I_\epsilon}$
surface.  Let us call
$\Sigma_{i_\epsilon}$ the surface containing the
intersection $i$ (at every fixed $\epsilon$), and $e_i$ the
edge through the intersection $i$.  Notice that
${\left\langle{{S}}\right|}\hat A^2_{\Sigma_{I_\epsilon}}$
vanishes for all surfaces $I_\epsilon$ except the ones
containing intersections. Thus the sum over surfaces
$\sum_{I_\epsilon}$ reduces to a sum over intersections.
Bringing the limit inside the sum and the square root, we
can write
\begin{eqnarray}
  {\left\langle{{S}}\right|} \hat A[\Sigma] &=& 
  \sum_{i\in\{S\cap\Sigma\}}\ {\left\langle{{S}}\right|} \sqrt{\hat 
A^2_i} 
\label{root2}\label{sumarea}
\\
\hat A^2_i &=& \lim_{\epsilon\to 0} \hat A^2_{i_\epsilon} 
\label{limit2}
\end{eqnarray}

For finite $\epsilon$, the state ${\left\langle{{S}}\right|}\hat
A^2_{i_\epsilon}$ has support on the union of the graphs
of $S$ and the graph of the loop $\alpha_{\sigma\tau}$ in
the argument of the operator (\ref{sigmatau}). But the
last converges to a point on $\Gamma_S$ as $\epsilon$ goes
to zero.  Therefore
\begin{equation}
	\lim_{\epsilon\to 0} \Gamma_{{\langle S |} 
           \hat A^2_{i_\epsilon}} 
	= \Gamma_S. 
\end{equation} 
The operator $\hat A[\Sigma]$ does not affect the graph of
${\langle S |}$.  Next, we have to compute the combinatorial
part of the action of the operator. 
  By equation (\ref{sumarea}), this is
given by a sum of terms, one for each
$i\in\{S\cap\Sigma\}$. Consider one of these terms.  By
definition of the ${\hat{\cal T}}$ loop operators and of the grasp
operation (Section 3), this is obtained by inserting two
trivalent intersections on the spin network edge $e_i$, 
connected by a new edge of color 2.
(The circle $\Gamma_{\alpha_{\sigma\tau}}$
has converged to a point on $e_i$; in turn, this point is
then graphically expanded 
following back and forward a segment connecting the two
intersections.  By indicating the representation of the
spin network simply by means of its $e_i$ edge, we thus
have
\begin{eqnarray}
{\left\langle{{\big|}^{p_e}~}\right|}~ 
   \hat A^2_{i_\epsilon} &=&\frac{1}{2}
          \int_{\Sigma_{i_\epsilon}\otimes\Sigma_{i_\epsilon}} 
          \!\!\! d^2\!\sigma  d^2\!\tau ~ 
          n_a(\sigma) n_b(\tau) 
          ~{\left\langle{{\big|}^{p_e}~}\right|}~
         {\hat{\cal T}}^{ab}[\alpha_{\sigma\tau}](\sigma,\tau) 
\\
&=& \!\!\!- \frac{l_0^4}{2} 
        \int_{\Sigma_{i_\epsilon}\otimes\Sigma_{i_\epsilon}} 
        \!\!\! d^2\!\sigma  d^2\!\tau ~ 
          n_a(\sigma)  \Delta^a[\beta_e,\sigma] 
          n_b(\tau)    \Delta^b[\beta_e,\tau]\  
          p_e^2 
{\left\langle{ \begin{array}{c}
  \setlength{\unitlength}{1 pt}
  \begin{picture}(30,40)
  \put(12, 0){\line(0,1){40}}
  \put(12,10){\circle*{4}}\put(12,30){\circle*{4}}
  \put( 0,4){${\scriptstyle  p_e}$}
  \put( 0,20){${\scriptstyle  p_e}$}
  \put( 0,34){${\scriptstyle  p_e}$}
  \put(12,20){\oval(20,20)[r]}
  \put(24,20){${}^2$}
\end{picture}\end{array} }\!\right|}
\nonumber
\end{eqnarray}
where we have already taken the limit (inside the
integral) in the state enclosed in the brackets
${\langle {~~} |}$. Notice that this does not depend on the
integration variables anymore, because the loop it
contains does not represent the grasped loop for a finite
$\epsilon$.
Notice also that the two integrals are independent, and
equal. Thus, we can write
\begin{equation}
~{\left\langle{{\big|}^{p_e}~}\right|}~ 
      \hat A^2_{i_\epsilon} = - \frac{l_0^4}{2} \left(
          \int_{\Sigma_{I_\epsilon}} 
          \!\!\! d^2\!\sigma   ~ 
          n_a(\sigma)  \Delta^a[\beta_e,\sigma] \right)^2
          p_e^2 
{\left\langle{ \begin{array}{c}
  \setlength{\unitlength}{1 pt}
  \begin{picture}(30,40)
  \put(12, 0){\line(0,1){40}}
  \put(12,10){\circle*{4}}\put(12,30){\circle*{4}}
  \put( 0,4){${\scriptstyle  p_e}$}
  \put( 0,20){${\scriptstyle  p_e}$}
  \put( 0,34){${\scriptstyle  p_e}$}
  \put(12,20){\oval(20,20)[r]}
  \put(24,20){${}^2$}
\end{picture}\end{array} }\right|} 
\end{equation}
The parenthesis is easy to compute. Using (\ref{delta}),
it becomes the analytic form of the intersection number
between the edge and the surface
\begin{eqnarray}
\int_{\Sigma_{i_\epsilon}} 
      \!\!\! d^2\!\sigma   ~ 
      n_a(\sigma)  \Delta^a[\beta_e,\sigma] 
&=& \int_{\Sigma_{i_\epsilon}} 
         \!\!\! d^2\!\sigma   ~ n_a(\sigma)  
         \int_{\beta_e} d\tau ~\dot{\beta}^a_e(\tau) 
         \delta^3 [\beta_e(\tau),s]
\nonumber \\
 &=& \pm 1,
\label{eq:IntersectionN}
\end{eqnarray}
where the sign, which depends on the relative orientation of
the loop and the surface, becomes then irrelevant because of
the square.  Thus
\begin{equation}
~{\left\langle{{\big|}^{p_e}~}\right|}~ \hat A^2_{i} 
= - \frac{l_0^4}{2}\  p_e^2 
{\left\langle{ \begin{array}{c}
  \setlength{\unitlength}{1 pt}
  \begin{picture}(30,40)
  \put(12, 0){\line(0,1){40}}
  \put(12,10){\circle*{4}}\put(12,30){\circle*{4}}
  \put( 0,4){${\scriptstyle  p_e}$}
  \put( 0,20){${\scriptstyle  p_e}$}
  \put( 0,34){${\scriptstyle  p_e}$}
  \put(12,20){\oval(20,20)[r]}
  \put(24,20){${}^2$}
\end{picture}\end{array} }\right|}  ,
\end{equation}
where we have trivially taken the limit (\ref{limit2}),
since there is no residual dependence on $\epsilon$. We
have now to express the tangle inside the bracket in terms
of (an edge of) a spin network state.  But tangles satisfy recoupling 
theory, and we can therefore
use the formula (E.8) in the appendix of \cite{dpr}, 
obtaining
\begin{eqnarray}
~{\left\langle{{\big|}^{p_e}~}\right|}~ \hat A^2_{i_\epsilon} &=& 
 - l_0^4~ p_e^2~ \frac{\theta(p_e,p_e,2)}{2 
       \Delta_{p_e}}~{\left\langle{{\big|}^{p_e}~}\right|}~ =  
 \nonumber \\
 = l_0^4~ \frac{p_e(p_e+2)}{4} ~{\left\langle{{\big|}^{p_e}~}\right|}~
&=& l_0^4~ \frac{p_e}{2}\left(\frac{p_e}{2}+1\right)
     ~{\left\langle{{\big|}^{p_e}~}\right|}. 
\nonumber
\end{eqnarray}
The square root in (\ref{root2}) is now easy to take
because the operator $\hat A^2_i$ is diagonal.
\begin{equation}
~{\left\langle{{\big|}^{p_e}~}\right|}~ \hat A_i 
  =~{\left\langle{{\big|}^{p_e}~}\right|}~ \sqrt{\hat A^2_i} = 
~=~ \sqrt{l_0^4~ \frac{p_e}{2}\left(\frac{p_e}{2}+1\right)}
     ~{\left\langle{{\big|}^{p_e}~}\right|}. 
\end{equation}
Inserting in the sum (\ref{root2}), we obtain the final result
\begin{equation}
{\langle S |}\ \hat{A}[\Sigma] =
\left({l^2_0\over 2} \sum_{i\in\{S\cap\Sigma\}}
\sqrt{p_i(p_i+2)}\right) \  {\langle S |}
\end{equation}
This result shows that the spin network states (with
a finite number of intersection points with the surface
and no vertices on the surface) are eigenstates of the
area operator. The corresponding spectrum is labeled by
multiplets  $\vec p = (p_1, ..., p_n)$ of positive half
integers, with arbitrary $n$, and given by
\begin{equation}
   A_{\vec p}\,[\Sigma] = {l^2_0\over 2} \, \sum_i  \sqrt{p_i(p_i+2)}.
\label{spec}
\end{equation}
Shifting from color to spin notation, we have 
\begin{equation}
   A_{\vec j}\,[\Sigma] = l^2_0 \, \sum_i  \sqrt{j_i(j_i+1)},
\end{equation}
where $j_1, ..., j_n$ are half integer. This expression reveals the 
$SU(2)$ origin of the spectrum. 

A similar result has been obtained for the volume. 

\section{Two applications}

The first hint on the thermodynamical behavior of black holes comes from
classical general relativity.  Hawking's theorem \cite{hawking} tells us
that the area of the event horizon of a black hole cannot decrease in
time, in classical general relativity.  In ref.\ \cite{bek}, Bekenstein
speculated that one can associate an entropy $S(A)$ to a 
Schwarzschild black hole of surface area $A$, where 
  \begin{equation}
                 S = c\ {k\over \hbar G} \ A
  \label{uno}
  \end{equation}
($c$ is a constant of the order of unity, $k$ the Boltzman constant,
and I put the speed of light equal to one). Bekenstein provided a number 
of physical arguments supporting this idea; but the reaction of the 
physicists community was cold, mainly due to the fact that since the 
black hole area $A$ is connected to the black hole energy $M$ by
\begin{equation}
	M = \sqrt{A \over 16\pi G^2},
\label{ma}
\end{equation}
the standard thermodynamical relation $T^{-1}=k\, dS/dE$ 
would imply the existence of a black hole temperature 
\begin{equation}
	T = {\hbar \over c 32 \pi k G M},
\end{equation}
and therefore in vacuum the black hole should emit thermal radiation at 
this temperature: a result difficult to believe. However, shortly after 
Bekenstein's suggestion, Hawking \cite{haw} derived black hole emission 
just from quantum field theory in curves spacetime. Hawking computed the 
emission temperature to be 
\begin{equation}
        T = {\hbar \over 8 \pi k G M},
\end{equation}
which beautifully supports Bekenstein's speculation, and fixes the 
constant $c$ at
\begin{equation}
        c_{Hawking} = {1\over 4}.
\end{equation}
Hawking's result opens many problems. I will consider two of these 
problems. First, in Hawking's derivation the quantum properties of 
gravity are neglected. Are these affecting the result?  Second, in 
general we  understand macroscopical entropy in statistical mechanical 
terms as an effect of microscopical degrees of freedom. What are the 
microscopical degrees of freedom responsible for (\ref{uno})? Can one 
derive (\ref{uno}) from first principles? Clearly a complete answer of  
these questions requires a quantum theory of gravity. 

\subsection{The Bekenstein-Mukhanov effect}

Recently, Bekenstein and Mukhanov \cite{bm} have suggested that the
thermal nature of Hawking's radiation may be affected by quantum
properties of gravity (For a review of earlier suggestions in this
direction, see \cite{lee}).   Bekenstein
and Mukhanov observe that in most approaches to quantum gravity the area
can take only quantized values \cite{garay}.  Since the area of the black
hole surface is connected to the black hole mass, black hole mass is
likely to be quantized as well.  The mass of the black hole decreases when
radiation is emitted.  Therefore emission happens when the black hole
makes a quantum leap from one quantized value of the mass (energy) to a
lower quantized value, very much as atoms do.  A consequence of this
picture is that radiation is emitted at quantized frequencies,
corresponding to the differences between energy levels.  Thus, quantum
gravity implies a discretized emission spectrum for the black hole
radiation. 

By itself, this result is not physically in contradiction with Hawking's
prediction of a continuous thermal spectrum. To understand this, consider
the black body radiation of a gas in a cavity, at high temperature.  This
radiation has a thermal Planckian emission spectrum, essentially
continuous.  However, radiation is emitted by elementary quantum emission
processes yielding a discrete spectrum.  The solution of the apparent
contradiction is that the spectral lines are so dense in the range of
frequencies of interest, that they give rise --effectively-- to a
continuous spectrum.  Does the same happen for a black hole? 

In order to answer this question, we need to know the energy spectrum of
the black hole, which is to say, the spectrum of the Area. Bekenstein and
Mukhanov pick up a simple ansatz: they assume that the Area is quantized
in multiple integers of an elementary area $A_0$.  Namely, that the area
can take the values 
 \begin{equation}
                        A_n=n A_0,
\label{ansatz}
\end{equation}
 where $n$ is a positive integer, and $A_0$ is an elementary area of the
order of the Planck Area 
\begin{equation}
                A_0=\alpha\hbar G,
\end{equation}
 where $\alpha$ is a number of the order of unity ($G$ is Newton's
constant and $c=1$).  Ansatz (\ref{ansatz}) is reasonable; it agrees, for
instance, with the partial results on eigenvalues of the area in the loop
representation given in \cite{weave}, and with the idea of a quantum
picture of a geometry made by elementary ``quanta of area''.  Since the
black hole mass is related to the area by (\ref{ma}), 
it follows from this relation and the ansatz (\ref{ansatz}) that the 
energy spectrum of the black hole is given by 
\begin{equation}
        M_n = \sqrt{{n \alpha \hbar \over 16\pi G }}.
\label{mass}
\end{equation}
 Consider an emission process in which the emitted energy is much smaller
than the mass $M$ of the black hole. From (\ref{mass}), the spacing
between the energy levels is
 \begin{equation}
        \Delta M =  {\alpha\hbar  \over 32\pi G M}.
\end{equation}
 From the quantum mechanical relation $E=\hbar\omega$ we conclude that
energy is emitted in frequencies that are integer multiple of the
fundamental emission frequency 
 \begin{equation}
                \bar\omega={\alpha \over 32\pi G M}. 
\end{equation}
 This is the fundamental emission frequency of Bekenstein and Mukhanov
\cite{bm} (they assume $\alpha=4\ln 2$).  Bekenstein and Mukhanov proceed
in \cite{bm} by showing that the emission amplitude remains the same as
the one in Hawking's thermal spectrum, so that the full emission spectrum
is given by spectral lines at frequencies multiple of $\bar\omega$, whose
envelope is Hawking's thermal spectrum. 

 As emphasized by Smolin in \cite{lee}, however, the Bekenstein-Mukhanov
spectrum is drastically different than the Hawking spectrum.  Indeed,
the maximum of the Planckian 
emission spectrum of Hawking's thermal radiation is around 
\begin{equation}
        \omega_H \sim {2.82 k T_H \over\hbar} = {2.82\over 8\pi GM} 
        = {2.82\cdot 4\over\alpha}\, \bar\omega \approx \bar\omega.
\end{equation}
 That is: the fundamental emission frequency $\bar\omega$ is of the same
order as the maximum of the Planck distribution of the emitted radiation.
It follows that there are only a few spectral lines in the regions where
emission is appreciable. Therefore the Bekenstein-Mukhanov spectrum is
drastically different than the Hawking spectrum: the two have the same
envelope, but while Hawking spectrum is continuous, the
Bekenstein-Mukhanov spectrum is formed by just a few lines in the interval
of frequencies where emission is appreciable.  Notice that such a
discretization of the emission spectrum is derived by Bekenstein and
Mukhanov on purely kinematical grounds, that is using only the (assumed)
spectral properties of the area.  To emphasize this fact, we will denote
it as the kinematical Bekenstein-Mukhanov effect. 

This result is of great interest because, in spite of its weakness, black
hole radiation is still much closer to the possibility of (indirect)
investigation than any quantum gravitational effect of which we can think.
Thus, a clear quantum gravitational signature on the Hawking spectrum is a
very interesting effect. Is this Bekenstein-Mukhanov effect credible?

As first suggested in \cite{lee}, and, independently, by Br\"ugmann, one
may use loop quantum gravity to check the Bekenstein-Mukhanov result, by
replacing the naive ansatz (\ref{ansatz}) with the precise spectrum
computed in loop quantum gravity. 

Consider a surface $\Sigma$ --in the present case, the event horizon of
the black hole--. The area of $\Sigma$ can take only a set of quantized
values. These quantized values are labeled by unordered n-tuples of
positive integers $\vec p = (p_1, ...  , p_n)$ of arbitrary length $n$. 
The spectrum is given in (\ref{spec}). If we disregard for a moment the
term $+1$ under the square root in (\ref{spec}), we obtain immediately the
ansatz (\ref{ansatz}), and thus the Bekenstein-Mukhanov result.  However,
the $+1$ is there.  Let us study the consequences of its presence.  First,
let us estimate the number of Area eigenvalues between the value $A>>>l_0$
and the value $A+dA$ of the Area, where we take $dA$ much smaller than $A$
but still much larger than $l_0$.  Since the $+1$ in (\ref{spec}) affects
in a considerable way only the terms with low $p_i$, we can neglect it for
a rough estimate. Thus, we must estimate the number of unordered strings
of integers $\vec p = (p_1, ... , p_n)$ such that 
\begin{equation}
        \sum_{i=1,n} p_i = {A\over 8\pi\hbar G}>>1.
\end{equation}
 This is a well known problem in number theory.  It is called the
partition problem. It is the problem of computing the number $N$ of ways
in which an integer $I$ can be written as a sum of other integers.  The
solution for large $I$ is a classic result by Hardy and Ramanujan
\cite{Andrews}. According to the Hardy-Ramanujan formula, $N$ grows as the
exponent of the square root of $I$. More precisely, we have for large $I$
that
 \begin{equation}
        N(I) \sim {1\over 4\sqrt{3}I} e^{\pi\sqrt{{2\over 3}I}}.
\end{equation}
Applying this result in our case we have that the number of 
eigenvalues between $A$ and $A+dA$ is
\begin{equation}
        \rho(A) \approx e^{\sqrt{\pi A\over 12\hbar G}}. 
\end{equation}
 Now, because of the presence of the $+1$ term, eigenvalues will overlap
only accidentally: generically all eigenvalues will be distinct. 
Therefore, the average spacing between eigenvalues decreases exponentially
with the inverse of the square of the area.  This result is to be
contrasted with the fact that this spacing is constant and of the order of
the Planck area in the case of the naive ansatz (\ref{ansatz}). This
conclusion empties the Bekenstein-Mukhanov argument.  Indeed,
the density of the energy levels becomes
 \begin{equation}
        \rho(M) \approx e^{\sqrt{4\pi G\over 3\hbar}M},
\label{density}
\end{equation}
 and therefore the spacing of the energy levels decreases {\it
exponentially\ } with $M$.  It follows that for a macroscopical black hole
the spacing between energy levels is infinitesimal, and thus the spectral
lines are virtually dense in frequency.  We effectively recover in this
way Hawking's thermal spectrum (except, of course, in the case of a Planck
scale black hole). A weaker but rigorous lower bound on the density of
eigenvalues, consistent with the argumented given here, is given in
\cite{area2}.  The conclusion is that the Bekenstein-Mukhanov effect
disappears if we replace the naive ansatz (\ref{ansatz}) with the spectrum
(\ref{spec}) computed from loop quantum gravity.  More generally, the
kinematical Bekenstein-Mukhanov effect is strongly dependent on the
peculiar form of the naive ansatz (\ref{ansatz}), and it is not robust. 
In a sense, this is a pity, because we loose a possible window on quantum
geometry. 

Mukhanov and, independently, Smolin have noticed that the possibility is
still open for the existence of a ``dynamical'' Bekenstein-Mukhanov
effect \cite{ms}. For instance, transitions in which a single Planck unit
of area is lost could be strongly favored by the dynamics.  To explore if
this is the case, one should make use of the full machinery of quantum
gravity, for instance by computing transition probabilities between
horizon's area eigenstates induced in a first order perturbation expansion
by the coupling between the area of the horizon and a surrounding
radiation field.  This could perhaps be done following the lines of
Ref.~\cite{jmp}. 

The conclusion is that the argument for the discretization of the black
hole emission spectrum given by Bekenstein and Mukhanov is not valid, if
we use quantitative result from loop quantum gravity.  As emphasized by
Mukhanov, this fact does not prove that the spectrum is indeed continuous,
since a discretization could be still be consequence of other (dynamical)
reasons. 

\subsection{Black Hole Entropy from Loop Quantum Gravity}

Finally, I present a derivation \cite{Rovelli96,kirill} of the
Bekenstein-Hawking expression (\ref{uno}) for the entropy of a
Schwarzschild black hole of surface area $A$ via a statistical mechanical
computation \cite{tutti}.  The strategy I follow is based on the idea that
the entropy of the hole originates from the microstates of the horizon
that correspond to a given macroscopic configuration. 

This idea was first suggested in a seminal work by York \cite{York}.  York
notices that the hole's radiance implies that the (macroscopic) event
horizon is located slightly inside the quasistatic timelike limit-surface,
leaving a thin shell between the two, which he proposes to interpret as
the region over which the microscopic horizon fluctuates. He interprets
these fluctuations as zero point quantum fluctuations of the horizon's
quasinormal modes, and, by identifying the thermal energy of these
oscillations with the shell's (``irreducible'') mass, he is able to
recover Hawking's temperature.  I take two essential ideas from York's
work: that the source of the hole entropy is in the degrees of freedom
associated with the fluctuations of the shape of the (microscopic)
horizon; and that the quasilocal measure of mass-energy governing
energetic exchanges between the horizon and its surroundings can be taken
as the Christodoulou-Ruffini \cite{ruffini} ``irreducible mass''
\begin{equation}
	M_{CR} = \sqrt{{A\over 16\pi G^2}}.
\label{CR}
\end{equation}
Can we replace York's perturbative semiclassical approach 
with a direct calculation within nonperturbative quantum gravity?

The relevance of horizon's surface degrees of freedom for the entropy has
been recently explored from various perspectives \cite{tf}.  (See also
\cite{maggiore} for an attempt to use the ``membrane paradigm''
\cite{membrane}: interactions of a black hole with its surroundings can be
described in terms of a fictitious physical membrane located close to the
horizon). An approach strictly related to the one I am going to describe
has been suggested in Refs.~\cite{carlip}, where it is argued that a
physical split of a gauge system gives rise to boundary degrees of
freedom, since the boundary breaks the gauge group. Using this idea the
Bekenstein-Hawking formula can be derived, by counting boundary states, in
3-d gravity. The relation is the following.  In GR, the broken component
of the gauge group includes diffeomorphisms that move the surface, and the
boundary degrees of freedom can probably be viewed as fluctuations of the
horizon. 

Consider a physical system containing a non-rotating and non-charged black
hole (say a collapsed star) as well as other physical components such as
dust, gas or radiation, which we denote collectively (improperly) as
``matter''.  We are interested in the statistical thermodynamics of such a
system.  A key observation is that because of Einstein's equations the
microscopic time-dependent inhomogeneities of the matter distribution
generate time-dependent ``microscopic'' inhomogeneities in the
gravitational field as well.  One usually safely  disregards these ripples 
of the
geometry.  For instance, we say that the geometry over the Earth's surface
is Minkowski (or Schwarzschild, due to the Earth gravitational field),
disregarding the inhomogeneous time-dependent gravitational field
generated by each individual fast moving air molecule. The Minkowski
geometry is therefore a ``macroscopic'' coarse-grained average of the
microscopic gravitational field surrounding us.  
However, in a statistical-thermodynamical treatment,
these fluctuations should not be disregarded, because they are precisely
the sources of the thermal behavior. 

Statistical thermodynamics is based on the distinction between the
macroscopic state of a system, determined by coarse-grained averaged
physical quantities, and its macroscopic state determined by a
(hypothetical) complete description of the system's dynamics.  A system in
equilibrium at a finite temperature $T$ is macroscopically stationary. 
However, its microstate fluctuates over microscopic non-stationary
configurations.  The family of the microstates over which the system
fluctuates when in a given macrostate form the statistical ``ensemble''
associated to the given macrostate.  For instance, the macrostate of a gas
in thermal equilibrium in a box is time-independent and spatially
homogeneous, while the microstates in the corresponding ensemble are
individually time dependent and non-homogeneous.  Thus, we must have
{\it two descriptions\ } of a physical black hole interacting with
surrounding matter at finite temperature.  The macroscopic
description is a stationary coarse grained description in which
inhomogeneities are smoothed out. The microscopic description does not 
neglect the minute thermal motions. 

Macroscopically, a non-charged and non-rotating hole is described by a
stationary metric with non-charged and non-rotating event horizon.  There
is only a one-parameter family of solutions of Einstein equations with
such properties: Schwarzschild with mass $M$, and corresponding 
event-horizon area $A=16\pi G^2M^2$.  Therefore in a thermal context the
Schwarzschild metric represents the coarse grained description of a
microscopically fluctuating geometry.  Microscopically the gravitational
field is non-stationary (because it interacts with non-stationary matter)
and non-spherically symmetric (because matter distribution is spherically
symmetric on average only, and not on individual microstates).  Its
microstate, therefore is {\it not\/} given by the Schwarzschild metric,
but by some complicated time-dependent non-symmetric metric.

I am convinced that taking such time-dependent non-symmetric microstates
of the geometry into account is essential for a statistical understanding
of the thermal behavior of black holes -- as it is in understanding the
thermal properties of any other system. Searching for a derivation of
black hole thermodynamics from properties of stationary or symmetric
metrics alone is like trying to derive the thermodynamics of an ideal gas
in a spherical box just from spherically symmetric motions of the
molecules. 

Thus, consider the microstate of our system.  Let us foliate spacetime
with a family of spacelike surfaces $\Sigma_t$, labeled by a time
coordinate $t$.  The intersection $h_t$ between the surface $\Sigma_t$ and
the future boundary of the past of future null-infinity defines the
instantaneous (microscopic) configuration of the event horizon at time
$t$.  Thus, $h_t$ is a closed 2-d surface immersed in $\Sigma_t$.  For
most times, this microscopic configuration of the event horizon is not
spherically symmetric.  Let us denote by $g_t$ the intrinsic and extrinsic
geometry of the horizon $h_t$.  Let $\cal M$ be the space of all possible
(intrinsic and extrinsic) geometries of a 2-d surface.  As $t$ changes,
the (microscopic) geometry of the horizon changes. Thus, $g_t$ wanders in
$\cal M$ as $t$ changes. 

I now recall some standard techniques in statistical mechanics in a form
that can be applied to our system.  Consider a thermodynamical system
$\cal S$, say an ideal gas in a isolated box.  Consider an equilibrium
macrostate of $\cal S$.  Under suitable ergodicity conditions, the
microstate of the system changes freely subjected to global conservation
laws only.  If the system is conservative and energy is the only conserved
quantity, then the system will wander in the entire region of its phase
space defined by a given total energy.  Next, we can ideally split $\cal
S$ into two subsystems ${\cal S}_1$ and ${\cal S}_2$, say two regions of
the box, separated by a thin film.  We are interested in studying the
thermal interactions between the two subsystems.  One approach is provided
by the microcanonical point of view.  Let us {\it ideally\ } isolate the
subsystem ${\cal S}_1$. Namely let us momentarily assume that it cannot
exchange heat.  Let $E_1$ be its energy, and $S_1(E_1)$ its entropy,
defined as the number of microstates that have energy $E_1$.  We now relax
the assumption that heat cannot be exchanged, and consider the full system
$\cal S$. If a small amount of heat $dQ$ is transferred from ${\cal S}_1$
to ${\cal S}_2$ the number of states available to ${\cal S}_1$ decreases
by an amount $(dS_1/dE_1)dQ$ and the number of microstates available to
${\cal S}_2$ increases by an amount $(dS_2/dE_2)dQ$.  The total number of
microstates available to $\cal S$ changes by \begin{equation} \delta N =
\left({dS_2\over dE_2}-{dS_1\over dE_1}\right)dQ.  \end{equation} From the
assumption that the equilibrium macroscopical configuration is the one to
which most microstates correspond, it follows that at equilibrium no small
heat transfer $dQ$ may increase the total number of available microstates,
and therefore 
 \begin{equation} 
	{dS_2\over dE_2}={dS_1\over dE_1}.
 \end{equation} 
Namely, the temperatures of the two systems are equal. 

Let us apply these ideas to our system. Consider our system as formed by
two sub-systems: the black hole and the rest.  We want to associate an
entropy $S$ to the black hole, where $S$ counts the number of microstates
over which the hole may fluctuate in an ideal situation in which no heat
(energy) is exchanged between the hole and its surroundings.  The precise
specification of this ensemble of microstates is crucial, and I now
discuss it in detail. 
 
First of all, as already noticed microscopic configurations do not need to
be individually spherically symmetric.  Second, only configurations of the
hole itself, and not the configurations of the surrounding geometry,
should affect the hole's entropy.  Thus, we must focus on the state of the
hole alone.  Next, we are considering the thermodynamic behavior of a
system containing the hole. This behavior cannot be affected by the hole's
interior.  The black hole interior may be in one of an infinite number of
states indistinguishable from the outside. For instance, the black hole
interior may (in principle) be given by a Kruskal spacetime; so that on
the other side of the hole there is another ``universe'' (say spatially
compact, if not for the hole) possibly with billions of galaxies.  This
potentially infinite number of such internal states does not affect the
interaction of the hole with its surroundings and is irrelevant here, 
because it cannot affect the energetic exchanges between the hole and the 
outside, which are the ones that determine the entropy.

Therefore we are only interested in configurations of the hole {\it that
have (microscopically) distinct effects on exterior of the hole}.  From
the exterior, the hole is completely determined by the geometrical
properties of its surface.  Thus, the entropy relevant for the
thermodynamical description of the thermal interaction of the hole with
its surroundings is entirely determined by the state of the gravitational
field (of the geometry) on the black hole surface, namely by $g_t$. 
 
Next, we have to determine the ``ensemble'' of the microstates $g_t$ over
which the hole may fluctuates under the ideal hypothesis of no heat
exchange.  In conventional statistical thermodynamics, one assumes that
the only conserved quantity is energy, and the microcanonical ensemble is
determined by fixing energy.  Here, however, there is no obvious candidate
for a notion of a conserved energy that could be used. 

A physical observation that leads us to the solution of this problem 
is that if energy  flows into the black hole then its area increases, 
while if the black hole  radiates away energy (via Hawking's radiation), 
then its area decreases.  Therefore we are lead to the idea that the 
(ideal) situation of no heat (energy) exchange is the evolution 
at fixed horizon's area.  Thus, following York, we take the 
Christodoulou-Ruffini quasi-local ``irreducible mass'' 
(\ref{CR}) as the relevant energy in this context (here $A$ is the area 
of $h_t$);  and we define the ensemble as the set of $g_t$ in $\cal M$ 
with the same $M_{CR}$, namely with the same area $A$. 

There is a number of reasons supporting the choice of this
ensemble. First, $M_{CR}$ is geometrically well defined, governs the
hole's energy exchanges, and agrees with the macroscopic black hole
energy. Second, the ensemble must contain reversible paths only.  In the
classical theory these conserve area (Hawking theorem \cite{hawking}).
Quantum theory allows classically forbidden energy exchanges with the
exterior (Hawking radiance), but it is unlikely, we believe, that it would
allow a nonreversible evolution of the horizon to become reversible
without energy exchange with the exterior. Third, we may reason backward
and let the thermodynamics indicate us the correct ensemble (which is how 
classical ensembles were first found). In this context, it perhaps
worthwhile recalling that difficulties to rigorously justifying a priori
the choice of the ensemble plague conventional thermodynamics anyway.

Summarizing, we are interested in counting the number $N(A)$ of states of
the geometry $g_t$ of a surface $h_t$ of area $A$, where different regions
of $h_t$ are distinguishable from each other.  The above discussion
indicates then that $S(A)=k \ln N(A)$ is the entropy we should associate
to the horizon in order to describe its thermal interactions with its
surroundings.  This ``number'' $N(A)$ meaningless in the classical theory. 
It is a this point only that we resort to the quantum theory.  As the
entropy of the electromagnetic field in a cavity is well defined only if
we take quantum theory into account, similarly we may expect that the
number of states $N(A)$ will be well defined in a quantum theory of
gravity.  The problem is thus to count the number of (orthogonal) quantum
states of the geometry of a two dimensional surface, having total area
$A$.  The problem is now well defined, and can be translated into a direct
computation.  

If a surface $\Sigma$ is given, its geometry is determined by its
intersections with the s-knot.  Intersections are of
three types: (a) an edge crosses the surface; (b) a vertex lies on the
surface; (c) a finite part of the s-knot lies on the
surface.  Intuitively, type (a) is the only ``generic'' case, and we
should disregard states of type (b) and (c). Ashtekar has suggested a
argument for neglecting type (b) and (c) intersections \cite{abhay}: we
wish to describe the geometry of a fluctuating surface $\Sigma$ as
observed from the exterior, and we expect the state of its geometry to be
stable under infinitesimal deformations of $\Sigma$. We may
thus consider the surface as the limit of a sequence of surfaces 
$\Sigma_\epsilon$, and its state as the (Hilbert norm) limit of the states
of $\Sigma_\epsilon$.  Clearly, states of type (b) and (c) cannot appear in
this way, and therefore we have to restrict our computation to states having
intersections of type (a) only \cite{roberto}. 
The quantum geometry on the surface is then determined by the {\it 
ordered\ } n-tuples of integers 
\begin{equation}
	\vec p=(p_1, ..., p_n) 
\end{equation}
that form the colors of the edges of type (a) intersections.

Notice that in the 
previous section we were interested in counting the density of the 
eigenvalues of the area (because these determine the density of the lines 
in the emission spectrum). While here we are interested in counting the 
density of the eigenstates. Thus, we must take the degeneracy of each 
eigenspace into account.  n-tuples that differ from each other in the 
ordering yield of course the same total area. Therefore they should be 
considered indistinguishable in counting eigenvalues.  On the other side 
they label distinct states. 

One may be tempted to observe that such states can be transformed 
into each other by diffeomorphisms, and therefore should not be 
considered  distinguishable. However, this observation is not correct. 
The point is that physical states are defined as equivalence classes under 
diffeomorphism of  the full space, not the surface alone. To understand 
this point, let us  consider a simplified analogy: Consider a set $A$, a 
set $B$, and a group $G$ that acts (freely) on $A$ and on $B$. Then $G$ 
acts on $A\times B$. What is  the space ${A\times B\over G}$? One may be 
tempted to say that it is  (isomorphic to) ${A\over G}\times{B\over G}$, 
but a moment of reflection  shows that this is not correct and the 
correct answer is 
\begin{equation}
{A\times B\over G} \sim {A\over G}\times B. 
\end{equation}
If $G$ does not act freely over $A$, we have to divide $B$ by the 
stability  groups of the elements of $A$.  Now, imagine that $A$ is the 
space of the  states of the exterior of the black hole, $B$ the space of 
the states of the black hole, and $G$ the diffeomorphism group of the 
horizon. Then we see that we must not divide $B$ by the 
diffeomorphisms of the surface, but only by those diffeomorphisms 
that leave the rest of the spin network invariant. As far as the state on 
the surface is concerned, this amounts to restrict to diffeomorphisms 
that do not mix the intersections between the spin network and the surface. 
Therefore n-tuples with different ordering must be considered as 
distinct.  Physically, this correspond to the fact that different 
locations in which the  spin network punctures the surface can be 
distinguished from each other in  terms of the external state of the 
gravitational field. For a more precise  version of these remarks, see 
\cite{donr}. 

Thus, our task is reduced to the task of counting the ordered n-tuples 
of integers $\vec p$ such that (\ref{area}). 
 More precisely, we are interested in the 
  number of microstates (n-tuples $\vec p$) such that the l.h.s of    
 (\ref{area}) is between $A$ and $A+dA$, where $A>>\hbar G$ and 
$dA$  is  much smaller than $A$, but still macroscopic. 
 
  Let   $    M = {A / 8\pi \hbar G }$,
  and let $N(M)$ be the number of ordered n-tuples $\vec p$, with 
  arbitrary $n$, such that
  \begin{equation}
            \sum_{i=1,n} \sqrt{p_i(p_i+2)} = M. 
  \label{part}
  \end{equation} 
  First, we over-estimate 
   $M(N)$ by approximating the l.h.s. of (\ref{part}) 
  dropping the $+2$ term under the square root. Thus, we want to   
 compute the number $N_+(M)$ of ordered n-tuples such that 
\begin{equation}
            \sum_{i=1,n} p_i = M. 
\label{piu}
  \end{equation}
  The problem is an exercise in combinatorics. It can 
  be solved, for instance, by noticing that if $(p_1, ..., p_n)$ 
is a partition of $M$ (that is, it solves (\ref{piu}) ), then  
$(p_1, ..., p_n,1)$ and $(p_1, ..., p_n+1)$ are partitions of 
$M+1$.  Since all partitions  of $M+1$ can be obtained in this
 manner, we have
\begin{equation}
            N_+(M+1)= 2 N_+(M). 
  \end{equation}
Therefore 
\begin{equation}
            N_+(M)= C\ 2^M. 
  \end{equation}
Where $C$ is a constant. In the limit of large $M$ we have
\begin{equation}
	\ln N_+(M) = (\ln 2)\ M. 
\end{equation}
Next, we under-estimate $M(N)$ by approximating (\ref{part}) as
\begin{equation}
	\sqrt{p_i(p_i+2)}=\sqrt{(p_i+1)^2-1} \approx (p_i+1).
\end{equation}
Thus, we wish to compute the number $N_-(M)$ of ordered n-tuples 
such 
that 
\begin{equation}
            \sum_{i=1,n} (p_i+1) = M. 
  \end{equation}
Namely, we have to count the partitions of $M$ in parts with 2 or 
more 
elements.  This problem can be solved by noticing that if $(p_1, ..., 
p_n)$ 
is one such partition of $M$ and $(q_1, ..., q_m)$ is one such 
partition 
of $M-1$, then $(p_1, ..., p_n+1)$ 
and $(q_1, ..., q_m,2)$ are partitions of $M+1$. All partitions of 
$M+1$ 
in parts with 2 or more elements can be obtained in this manner, 
therefore
\begin{equation}
            N_-(M+1)= N_-(M) +  N_-(M-1)  .
\label{mmm}
  \end{equation}
It follows that  
\begin{equation}
            N_-(M)= D a_+^M + E a_-^M
\label{apm} 
  \end{equation}
where $D$ and $E$ are constants and $a_\pm$ (obtained by inserting 
(\ref{apm}) in (\ref{mmm})) are  the two roots of the equation
\begin{equation}
	a_\pm^2=a_\pm+1.
\end{equation}
In the limit of large $M$ the term with the highest root dominates, 
and 
we have
\begin{equation}
	\ln N_-(M) = (\ln a_+)\ M = \ln{1+\sqrt{5}\over 2}\ M. 
\end{equation}
By combining the information from the two estimates, we conclude 
that 
\begin{equation}
	\ln N(M) = d\ M. 
\end{equation}
where
\begin{equation}
	  \ln {1+\sqrt{5}\over 2}  < d  < \ln 2  
\end{equation}
or
\begin{equation}
	0.48 <  d  < 0.69. 
\end{equation}
 Since the integers $M$ are equally spaced, our 
computation yields 
immediately the density of microstates. The 
number $N(A)$ of microstates with area 
$A$ grows for large $A$ as
  \begin{equation}
            \ln N(A) = d\  {{A\over8\pi \hbar G}}
\label{n}
  \end{equation}
 This gives immediately
  \begin{equation}
            S(A) =  c \ {k\over \hbar G} \  A . 
  \end{equation}
  which is the Bekenstein-Hawking formula.  For a different (and very
elegant) derivation, see \cite{kirill}.
  The constant of proportionality that we have 
  obtained is 
  \begin{equation}
            c = {d \over 8 \pi }, 
  \end{equation}
  which is roughly $4 \pi$ times smaller than Hawking's value    
  $c_{Hawking}={1\over  4}$. 
    
In summary: I have argued that the black hole entropy 
relevant for the hole's thermodynamical interaction with its 
surroundings is the number of the quantum 
microstates of the hole which have microscopically distinct 
effects on the exterior of the hole. I have argued that these
states are given by the quantum state of the horizon with the same area. 
 I have counted such microstates 
using loop quantum gravity. I have obtained that the entropy is 
proportional to the area, as in the Bekenstein-Hawking 
formula. 

Several issues remain open. I have worked in the simplified setting of a
hole interacting with a given geometry, instead of working within a fully
generally covariant statistical mechanics \cite{stat}. Also, it would be
nice to have a direct characterization of the event horizon in the quantum
theory: this could perhaps be given along the following lines. Consider a 
weave \cite{weave} state $|w\rangle$ which solves the hamiltonian 
constraint and represents a physical black hole. 
This can be expanded in the s-knot basis \begin{equation}
	|w\rangle = \sum_i c_i |s_i\rangle.
\end{equation}
Consider the observables $\hat O_j$ representing measurement at future 
null infinity (for instance, see \cite{fuzzy}). For every $s_i$, and all 
$\hat O_j$, define as ``internal'' the edges $l_k$ of $s_i$ such that the 
expectation values
\begin{equation}
	\bar O_j = \langle w|\hat O_j|w\rangle
\end{equation}
satisfy
\begin{equation}
	{d \bar O_j \over d l_k} = 0 ,
\end{equation}
meaning that $\bar O_j$ is not affected if we change the color of $l_k$. 
A similar definition can be given for ``internal vertices''. Denote edges 
and vertices that are not internal as external. Now the 
quantum event horizon can be defined as the set of external edges that 
are nor surrounded by external edges or external vertices only.  Clearly 
this captures the idea of the boundary between the region that 
``affects future null infinity'' and the regions that doesn't. Notice 
that under this definition the quantum event horizon is just a collection 
of edges (pictorially: the edges cut by the horizon). 
This approach might clarify the issue of the type (b) and
(c) intersections, and, I believe, deserves to be investigated.   

Finally, the numerical discrepancy with the Hawking's
value indicates that something is still poorly understood. Jacobson
\cite{ted} has suggested that finite renormalization effects of the Newton
constant might account for this discrepancy and has begun to explore
how the presence of matter might affect it.

\end{document}